\documentclass[%
 reprint,
superscriptaddress,
nofootinbib,
 amsmath,amssymb,
 aps,
pra,
]{revtex4-2}

\usepackage{graphicx}
\usepackage{dcolumn}
\usepackage{bm}
\usepackage{physics}
\usepackage{bbm}
\usepackage[hidelinks]{hyperref}
\makeatletter
\providecommand{\href@noop}[2]{#2}
\makeatother
\usepackage{cleveref}
\usepackage[title]{appendix}
\Crefformat{figure}{#2Fig.~#1#3}
\Crefmultiformat{figure}{Figs.~#2#1#3}{ and~#2#1#3}{, #2#1#3}{ and~#2#1#3}

\begin{document}

\preprint{APS/123-QED}

\title{Towards stable and accurate electron dynamics via neural network based time-dependent variational Monte Carlo}

\date{\today}

\author{Weizhong Fu}
\email{fuwz@pku.org.cn}
\affiliation{ByteDance Seed}
\affiliation{School of Physics, Peking University, Beijing 100871, People’s Republic of China}

\author{Zhe Li}
\affiliation{ByteDance Seed}

\author{Yubing Qian}
\affiliation{ByteDance Seed}
\affiliation{School of Physics, Peking University, Beijing 100871, People’s Republic of China}

\author{Ruichen Li}
\affiliation{ByteDance Seed}

\author{Weiluo Ren}
\email{renweiluo@bytedance.com}
\affiliation{ByteDance Seed}

\author{Ji Chen}
\email{ji.chen@pku.edu.cn}
\affiliation{School of Physics, Peking University, Beijing 100871, People’s Republic of China}
\affiliation{Interdisciplinary Institute of Light-Element Quantum Materials and Research Center for Light-Element Advanced Materials, Peking University, Beijing 100871, People’s Republic of China}
\affiliation{State Key Laboratory of Artificial Microstructure and Mesoscopic Physics and Frontiers Science Center for Nano-Optoelectronics, Peking University, Beijing 100871, People's Republic of China}

\begin{abstract}
Real-time dynamics of interacting electrons lies at the interface between quantum mechanics and non-equilibrium physics, governing the microscopic origin of ultrafast phenomena of molecules and nano-materials. 
%
Though neural network variational Monte Carlo has achieved unprecedented accuracy for stationary state calculations, its extension to real-time evolution remains challenging.
In this work, we introduce the neural basis time-dependent variational Monte Carlo framework, which achieves stable and highly accurate simulations of electron dynamics.
By constraining the time evolution to a compact, customized manifold spanned by the neural basis, we effectively bypass instability issues and achieve long-term stable evolution.
Moreover, we demonstrate that this framework yields benchmark-quality accuracy in simulating the laser-driven dipole responses of the hydrogen atom and a stretched hydrogen molecule, and accurately extracts the dynamic polarizabilities of helium and beryllium atoms.
Our work reveals the vast potential of neural network wavefunctions for accurately describing real-time electron dynamics and establishes a promising new route for first-principles simulations of complex, time-dependent electronic phenomena.
\end{abstract}

\maketitle



Electron dynamics is a critical problem at the frontier of modern physics and chemistry, underlying a wide variety of fundamental processes, including chemical reactivity, charge migration, and energy dissipation~\cite{zewail2000femtochemistry,cederbaum1999ultrafast,nisoli2017attosecond}.
Essentially, electron dynamics determines many macroscopic properties and responses of matter, such as induced dipole moments, dynamic polarizabilities, ionization, and multiphoton resonances~\cite{orr1971perturbation,brabec2000intense}.
From the theoretical point of view, electron dynamics occurs on an extremely fast timescale, often from femtoseconds to attoseconds, and is frequently accompanied by non-equilibrium phenomena induced by strong-field excitations.
These features, along with complex electron-electron correlation, make electron dynamics extraordinarily difficult to track, model and predict accurately.
Furthermore, with the rapid development of ultrafast experimental techniques~\cite{corkum2007attosecond,krausz2009attosecond}, the demand for solving many-body electron dynamics problems has become ever more urgent.

%
%

Traditional computational approaches for electron dynamics face an insurmountable accuracy-efficiency dilemma.
%
%
The widely employed time-dependent density functional theory (TDDFT) based methods are limited by the approximate exchange-correlation functionals and adiabatic approximations commonly used in practice~\cite{runge1984tddft,ullrich2012tddft}, with well-known failures for long-range charge transfer, double excitations, and strongly driven charge dynamics~\cite{dreuw2004failure,maitra2004double,fuks2013dynamics}.
%
Wavefunction methods based on deterministic solutions, such as real-time time-dependent configuration interaction and coupled-cluster theories, provide systematically improvable accuracy, but their computational cost grows steeply with system size and basis-set dimension~\cite{szabo1996qc,kvaal2012tdcc}.
In the past few years, neural network wavefunctions have emerged as a powerful tool for stationary state calculations, demonstrating unprecedented accuracy across various molecules and materials~\cite{carleo2017solving,pfau2020, choo2020fermionic, hermann2020deep, von2022self, lin2023explicitly, gerard2022gold, pescia2022neural, li2022, ren2023towards, fu2024variance, li2024computational,fu2025local,jiang2025neural,li2025spin,hermann2023ab,tang2025deep}.
Unlike deterministic expansions over a rapidly growing configuration space, neural network variational Monte Carlo uses compact nonlinear functions of electronic coordinates to encode many-body correlations and evaluates observables by stochastic sampling.
This combination of expressiveness and variational flexibility makes the extension of neural network wavefunctions to real-time evolution a promising avenue for tackling complex electron dynamics problems.
%
%
This extension, however, turns out to be a highly non-trivial task.
Nys et al. pioneered the integration of time-dependent variational Monte Carlo (tVMC)~\cite{carleo2017solving} with a neural network parameterized Slater-Jastrow-backflow wavefunction ansatz for real molecules, but success has not yet been achieved with more expressive neural network wavefunctions~\cite{nys2024ab}.  
%
In a different attempt, Hou et al. proposed a global spacetime optimization framework to obtain the time evolution trajectory of neural network wavefunctions~\cite{hou2025global}, but they find the performance under relatively strong electric fields remains unsatisfactory.
These valuable studies have motivated a better integration of neural network wavefunctions with tVMC, and also raise a pressing question of whether such wavefunctions can meet their expectation to address many-body electron dynamics problems.

\begin{figure}[htbp]
  \centering
  \includegraphics[width=86mm]{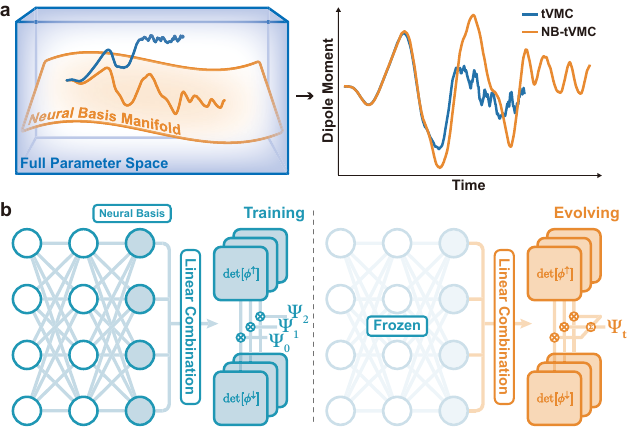}
  \caption{\textbf{(a)} Schematic comparison between full-parameter tVMC and NB-tVMC. Directly adopting tVMC to evolve in the full parameter space faces severe instability problems, while NB-tVMC maintains a long-term stable simulation.
  \textbf{(b)} Overview of the NB-tVMC framework. During multistate training, determinant blocks are assigned to individual low-lying eigenstates so that the shared nonlinear layers learn features relevant to the targeted spectrum. During real-time propagation, these features are frozen, and all determinants are recombined into a single time-dependent state evolving by updating the final linear layer.}
  \label{fig:frame}
\end{figure}

In this work, we introduce a neural basis time-dependent variational Monte Carlo (NB-tVMC) framework that achieves stable and highly accurate real-time evolution, as illustrated in \Cref{fig:frame}a.
The key idea is to first construct a compact, customized manifold of many-body states using a well-trained neural network wavefunction that effectively captures the ground and low-lying excited states.
The dynamics are then confined to this fixed manifold, where only a small set of linear coefficients are evolved.
We demonstrate that this approach yields results consistent with TDDFT and full configuration interaction (FCI) methods, and is superior to previous neural network evolution schemes under relatively strong electric fields.

Electron dynamics is governed by the time-dependent Schr\"{o}dinger equation~\cite{schrodinger1926wave},
\begin{equation}
i\frac{\partial}{\partial t}|\Psi(t)\rangle=\hat{H}(t)|\Psi(t)\rangle.
\end{equation}
For a parameterized ansatz $|\Psi(\boldsymbol{\theta}(t))\rangle$ with time-dependent parameters $\boldsymbol{\theta}(t) = \{\theta_1(t), \ldots, \theta_{N_p}(t)\}$, the problem reduces to determining $\boldsymbol{\theta}(t)$ such that $|\Psi(\boldsymbol{\theta}(t))\rangle$ approximates the exact evolution state $|\Psi(t)\rangle$ as closely as possible at each time $t$.
%

We represent the many-body electronic state, following the FermiNet architecture~\cite{pfau2020}, as the summation of products of spin-resolved determinants,
\begin{equation}
  \Psi = \sum_{k=1}^{N_{\mathrm{det}}} \det \left[\phi_{ij}^{k\uparrow}\right] \det \left[\phi_{ij}^{k\downarrow}\right],
\end{equation}
where determinant elements $\phi_{ij}^{k\alpha} = \phi_{i}^{k\alpha}(\mathbf{r}_j^{\alpha};\{\mathbf{r}_{/j}^{\alpha}\};\{\mathbf{r}^{\bar{\alpha}}\})$ ($\alpha$ is the spin index) are neural ``orbitals'' that implicitly depend on all electron coordinates.
%
These ``orbitals'' are constructed via a linear combination of the neural basis,
\begin{equation}
  \phi_{ij}^{k\alpha} = \sum_{\mu} \tilde{\theta}_{i\mu}^{k\alpha} \chi_{\mu j}^{\alpha}.
\end{equation}
Here, $\tilde{\theta}_{i\mu}^{k\alpha}$ are the parameters of the final linear layer of the network, while the neural basis functions $\chi_{\mu j}^{\alpha} = h_{\mu}^{L\alpha}(\mathbf{r}_j^{\alpha};\{\mathbf{r}_{/j}^{\alpha}\};\{\mathbf{r}^{\bar{\alpha}}\})$ are taken directly as the outputs of the last hidden layer $\mathbf{h}^L$.
These neural basis functions differ from fixed one-electron orbitals. 
Through the preceding nonlinear layers, they depend on the full electronic configuration and already encode electron-electron correlation before the final linear mixing.

As illustrated in \Cref{fig:frame}b, our NB-tVMC framework consists of two stages.
In the training stage, the neural network wavefunction is optimized to simultaneously represent multiple low-lying energy eigenstates using the NES-VMC procedure at $t = 0$~\cite{pfau2024excitedstates}.
Note that FermiNet and NES-VMC are specific choices in this work, and the same construction can be combined with other differentiable neural network wavefunctions~\cite{von2022self,li2024computational} and multistate training schemes that provide a suitable neural basis~\cite{li2024spin,szabo2024improved,schatzle2025ab,gao2026excited}.
In the evolving stage, $\tilde{\theta}_{i\mu}^{k\alpha} = \tilde{\theta}_{i\mu}^{k\alpha}(t)$ are the only variational parameters propagated by the tVMC method (details provided in Supplementary Note 2).
Equivalently, the evolution is restricted to the manifold spanned by the neural basis $\chi_{\mu j}^{\alpha}$, which is fixed after the training stage.
As shown in \Cref{fig:frame}b, once the training converges, this manifold accurately captures the target quantum states, including the ground state (typically used as the initial state) and low-lying excited states that play crucial roles in electron dynamics.
As displayed in the right panel of \Cref{fig:frame}a, NB-tVMC yields stable evolution trajectories, whereas directly applying tVMC to all network parameters can lead to severe instability issues and a rapid breakdown of the trajectory (see thorough discussions in Supplementary Note 3).
In the remainder of this paper, we show that despite evolving within such a restricted manifold, NB-tVMC produces accurate real-time dynamics across a range of systems, enabled by the customized manifold construction in the training stage.


\begin{figure*}[t]
  \centering
  \includegraphics[width=1.0\textwidth]{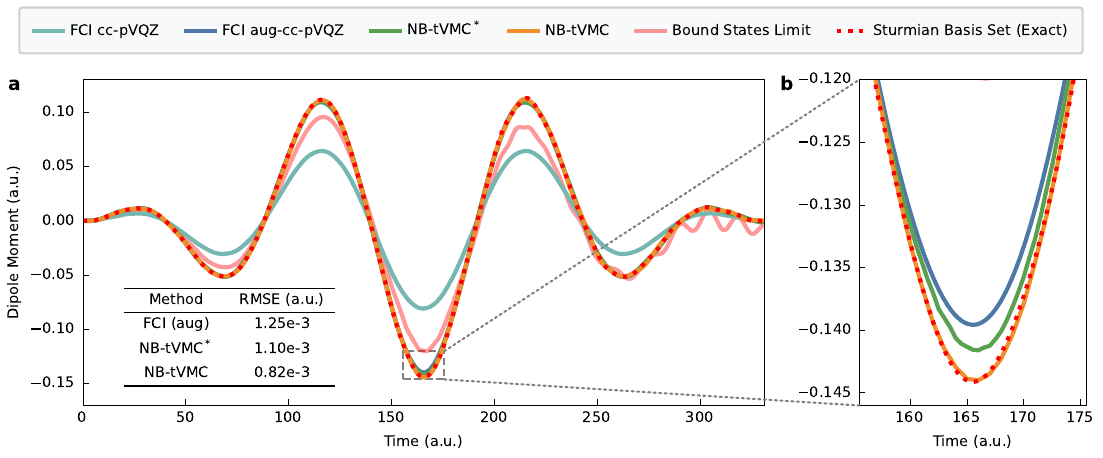}
  \caption{\textbf{(a)} Dipole moment evolution trajectories for a hydrogen atom under an external electric field using various methods. NB-tVMC$^*$ denotes a neural basis trained by targeting only the ground state. The pink curve is converged with a sufficiently large set of analytic bound eigenstates, while the red dashed curve is converged with a sufficiently large Sturmian basis and serves as the benchmark. The inset table reports the root-mean-square errors of the three most accurate results. \textbf{(b)} Zoom-in of the highest-peak region.}
  \label{fig:H_atom}
\end{figure*}

A common scenario in the study of electron dynamics is to investigate the real-time evolution of an electronic wavefunction driven by a varying electric field $\mathbf{E}(t)$.
Under the dipole approximation~\cite{cohentannoudji1992atomphoton}, the corresponding time-dependent Hamiltonian reads
\begin{equation}
  \hat{H}(t) = \hat{H}_0 - \hat{\boldsymbol{\mu}} \cdot \mathbf{E}(t),
\end{equation}
where $\hat{H}_0$ is the field-free Hamiltonian and $\hat{\boldsymbol{\mu}}$ is the total dipole moment operator.
To prevent unphysical high-frequency oscillations associated with an instantaneous switch-on of the external electric field at $t=0$, the field is typically modulated by an envelope function.
%
We first benchmark our method on a hydrogen atom subjected to an external laser pulse with a sine-squared envelope~\cite{han2010comparison}.
%
%
It is well established that the Sturmian basis yields near-exact electron dynamics in this scenario~\cite{rotenberg1970theory,dorr1990atomic,buchleitner1995microwave}, and it shows excellent agreement with the grid-based calculations in Ref.~\cite{han2010comparison}.
We therefore regard it as a reliable benchmark here, plotted as the red dashed curve in \Cref{fig:H_atom}.
This electric field is strong enough to induce contributions from Rydberg and even scattering states.
To demonstrate this, we display the evolution obtained from a converged set of analytic bound eigenstates, and it exhibits a clear deviation from the benchmark.
For the same reason, the FCI calculation using the cc-pVQZ basis performs poorly, whereas adding diffuse functions (aug-cc-pVQZ) substantially improves the accuracy.
Remarkably, our NB-tVMC method produces results that agree closely with the benchmark.
%
%
To assess the role of excited-state targets in the training stage, we also consider a ground-state-only variant, denoted as NB-tVMC$^*$.
As shown in \Cref{fig:H_atom}b, NB-tVMC$^*$ already outperforms FCI with the aug-cc-pVQZ basis, while expanding the training targets to include excited states further brings NB-tVMC into great agreement with the benchmark.
The benefit of such multistate training becomes even more evident in the density evolution, which we will discuss later.
Quantitatively, the root-mean-square errors over the full trajectory are $1.25\times10^{-3}$~a.u. for FCI/aug-cc-pVQZ, $1.10\times10^{-3}$~a.u. for NB-tVMC$^*$, and $0.82\times10^{-3}$~a.u. for NB-tVMC.
Achieving such accuracy in real-time propagation is particularly remarkable for a stochastic method, given that statistical noise may accumulate over time.
%
Furthermore, the close agreement of NB-tVMC with the Sturmian benchmark indicates that, although only low-lying levels are explicitly targeted during training, the optimized neural basis retains enough flexibility to represent the higher-energy and continuum-like components required by the subsequent dynamics.

\begin{figure*}[t]
  \centering
  \includegraphics[width=1.0\textwidth]{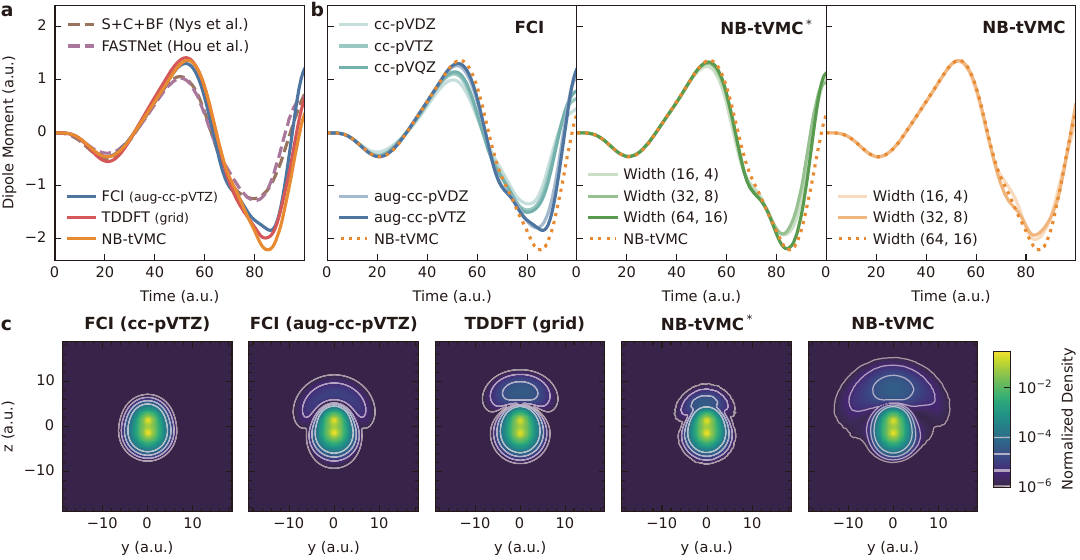}
  \caption{\textbf{(a)} Dipole moment evolution trajectories for a stretched H$_2$ molecule under an external electric field using various methods. \textbf{(b)} Convergence analysis of systematically improvable methods: FCI with increasing basis set size and NB-tVMC with increasing network width, where $(a, b)$ denotes the width of one-electron and two-electron streams, respectively. NB-tVMC$^*$ denotes a neural basis trained by targeting only the ground state. \textbf{(c)} Normalized electron density on the $y$--$z$ plane ($x=0$) at $t=67$~a.u. from FCI, grid-based TDDFT, and NB-tVMC calculations, illustrating their ability to represent the field-induced delocalized wave packet. The contour lines correspond to normalized densities ranging from $10^{-6}$ to $10^{-4}$.}
  \label{fig:H2}
\end{figure*}
We next study a stretched H$_2$ molecule driven by a laser pulse with a trapezoidal envelope.
This system was first investigated using the time-dependent Hartree-Fock method~\cite{li2005time} and has recently been revisited by methods based on neural network wavefunctions~\cite{nys2024ab,hou2025global}.
To rigorously test the capability of neural network wavefunctions, the H$_2$ molecule was stretched to twice its equilibrium bond length~\cite{nys2024ab}, which enhances electron correlation and also makes the system more susceptible to ionization, thereby presenting a more challenging problem.
Our NB-tVMC result under this circumstance is shown in \Cref{fig:H2}a, together with previous neural network calculations labeled as S+C+BF~\cite{nys2024ab} and FASTNet~\cite{hou2025global}, as well as FCI and TDDFT reference results (here we focus on the first three peaks before the weak ionization strongly affects the later-time trajectories, and the full time trajectories are provided in Supplementary Note 8).
%
%
%
%
Both S+C+BF and FASTNet substantially underestimate the peak heights relative to the FCI and TDDFT references, whereas NB-tVMC shows excellent agreement with them.
%
%
In \Cref{fig:H2}b, we further analyze the convergence trends of FCI results with increasing basis set size and NB-tVMC results with increasing network width.
As an additional check, we also include another group of NB-tVMC$^*$ results.
%
%
Across all cases, increasing either the basis set size or the network width systematically drives both the FCI and NB-tVMC$^*$ trajectories toward the highest-capacity NB-tVMC result (orange dashed curve).
%
Moreover, NB-tVMC trajectories with different network widths remain perfectly self-consistent up to $t = 70$ a.u.
All these observations suggest that our NB-tVMC method provides a benchmark-quality description for this challenging system.
It is also worth noting that FCI with the smaller aug-cc-pVDZ basis outperforms the larger cc-pVQZ basis.
This comparison suggests that a purely size-driven basis set enlargement, without guidance from physical insights, can be insufficient for improving real-time electron dynamics.
%
In contrast, the accuracy of NB-tVMC can be improved in a controlled manner by scaling the network capacity and by explicitly incorporating physically relevant target states during training.

%
%
%
%
%
%

Under this relatively strong driving field, the electronic wavefunction develops a spatially delocalized wave packet during real-time evolution.
To examine how different methods capture this feature, we compare in \Cref{fig:H2}c the normalized electron density on the $x=0$ plane at $t=67$~a.u.
The FCI result with the cc-pVTZ basis remains largely localized around the nuclei and therefore misses this delocalized component, while adding diffuse functions in FCI/aug-cc-pVTZ partially restores the spatial extension.
%
%
The grid-based TDDFT calculation, which represents the electronic state directly in real space, provides a more natural description of the density and clearly captures the delocalized wave packet.
Although NB-tVMC, like FCI, propagates within a parameterized manifold, the expressive neural basis captures the delocalization without explicitly adding diffuse functions.
While the NB-tVMC$^*$ result shows only a weakly extended wave packet, the multistate-trained NB-tVMC result yields prounounced delocalized density, demonstrating its strong expressiveness over a broad region of real space.
For truly ionizing regimes, however, we expect that special treatments such as complex absorbing potentials will still be required~\cite{muga2004complex}.

\begin{figure}[t]
  \centering
  \includegraphics[width=86mm]{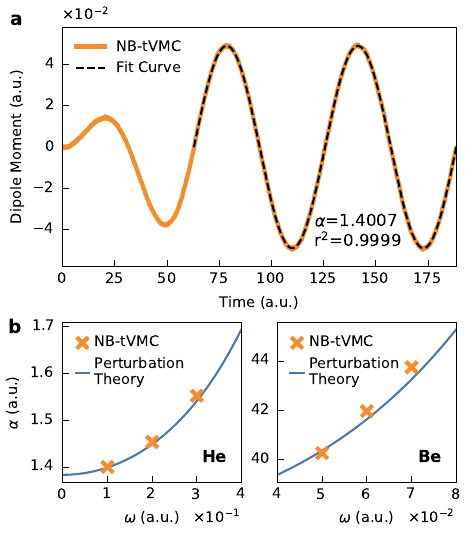}
  \caption{\textbf{(a)} Dipole moment response of the He atom to a monochromatic electric field ($\omega=0.1$~a.u.) and the corresponding dynamic polarizability fitting result. \textbf{(b)} Dynamic polarizability results of He and Be atoms at various driving frequencies. The curves labeled ``Perturbation Theory'' are literature results from calculations based on perturbation theory~\cite{puchalski2016theoretical,komasa2001dipole}. Additional fitting details are provided in Supplementary Note 9.}
  \label{fig:dp}
\end{figure}
Having demonstrated the accuracy and stability of our NB-tVMC method, we now apply it to calculate the dynamic polarizability $\alpha(\omega)$, which characterizes the linear response of a system's dipole moment to an oscillating external electric field of frequency $\omega$.
Specifically, for a monochromatic electric field $\mathbf{E}(t) = E_{\mathrm{max}} \sin(\omega t)\mathbf{\hat{z}}$ polarized along the $z$-direction, the induced dipole moment can be expanded as~\cite{orr1971perturbation}:
\begin{equation}
\mu_z(t) = \mu_z(0) + \alpha(\omega) \sin(\omega t)E_{\mathrm{max}} + \mathcal{O}(E_{\mathrm{max}}^2).
\end{equation}
In the weak-field regime, the oscillation amplitude of $\mu_z(t)$ scales linearly with the electric field strength, enabling a direct extraction of $\alpha(\omega)$ by fitting the $\mu_z(t)$ trajectory~\cite{ding2013efficient}.
%
%
Viewed in this light, the tendency of prior inaccurate evolution trajectories to underestimate peak amplitudes can be interpreted, at least in part, as an underestimation of the dynamic polarizability, which can be understood from a sum-over-states perspective (see Supplementary Note 4).

As an illustrative example, \Cref{fig:dp}a shows the NB-tVMC dipole trajectory for He at $\omega=0.1$~a.u.
The field amplitude is set to $E_{\mathrm{max}}=0.035$~a.u. and is ramped on with a linear envelope over one optical cycle.
The fit is then performed over the subsequent steady-state oscillation.
The dipole exhibits an essentially perfect sinusoidal response, further confirming the numerical stability of the NB-tVMC method.
The extracted dynamic polarizabilities over a range of frequencies are shown in \Cref{fig:dp}b, including results for the computationally more challenging Be atom.
The corresponding fitting details are provided in Supplementary Note 9.
Overall, our results agree closely with the perturbation-theory curves.
Since accurate $\alpha(\omega)$ across frequencies requires a reliable description of both excitation energies and transition dipole moments~\cite{orr1971perturbation}, this agreement indicates that NB-tVMC captures both ingredients with high fidelity.

In conclusion, we have introduced the NB-tVMC framework for stable and accurate real-time electron dynamics with expressive real-space neural network wavefunctions.
By constraining the time evolution to a customized compact manifold spanned by the neural basis, we overcome instability issues and achieve stable, long-time evolution trajectories.
Despite this constrained manifold, our method demonstrates exceptional accuracy, even under relatively strong electric fields.
The resulting dipole moment trajectories substantially improve upon those of other recent neural network-based methods and can even surpass traditional high-accuracy methods in the benchmark system, such as FCI with large basis sets.
In this work, we primarily employed NES-VMC to target several low-lying energy levels for the construction of a customized neural basis.
Looking ahead, the flexibility of NB-tVMC naturally allows for alternative choices of training targets, such as symmetry-resolved states, states optimized for specific observables, or eigenstates in the presence of strong static fields, which may possibly further improve performance in challenging regimes.
%
%
Furthermore, in conjunction with recently proposed neural network wavefunction architectures capable of handling particles beyond electrons, such as nuclei and muons~\cite{carr2026neural,cai2026permutation}, our method can be readily extended beyond pure electron dynamics.
NB-tVMC thus opens a practical route to first-principles real-time simulations with expressive neural network wavefunctions, which holds potential for broad applications in quantum dynamics.

\begin{acknowledgments}
We thank Han Wang and Enze Hou for fruitful discussions. 
This work was supported by the National Science Foundation of China under Grant No. 12334003.
\end{acknowledgments}


\bibliography{ref}

@article{nys2024ab,
  title={Ab-initio variational wave functions for the time-dependent many-electron Schr{\"o}dinger equation},
  author={Nys, Jannes and Pescia, Gabriel and Sinibaldi, Alessandro and Carleo, Giuseppe},
  journal={Nature communications},
  volume={15},
  number={1},
  pages={9404},
  year={2024},
  publisher={Nature Publishing Group UK London}
}

@article{hou2025global,
  title={A Global Spacetime Optimization Approach to the Real-Space Time-Dependent Schr\"{o}dinger Equation},
  author={Hou, Enze and Liu, Yuzhi and Wang, Lei and Wang, Han},
  journal={arXiv preprint arXiv:2511.12983},
  year={2025}
}

@article{han2010comparison,
  title={Comparison between length and velocity gauges in quantum simulations of high-order harmonic generation},
  author={Han, Yong-Chang and Madsen, Lars Bojer},
  journal={Physical Review A—Atomic, Molecular, and Optical Physics},
  volume={81},
  number={6},
  pages={063430},
  year={2010},
  publisher={APS}
}

@incollection{rotenberg1970theory,
  title={Theory and application of Sturmian functions},
  author={Rotenberg, Manuel},
  booktitle={Advances in atomic and molecular physics},
  volume={6},
  pages={233--268},
  year={1970},
  publisher={Elsevier}
}

@article{dorr1990atomic,
  title={Atomic hydrogen irradiated by a strong laser field: Sturmian basis calculations of rates for high-order multiphoton ionization, Raman scattering, and harmonic generation},
  author={D{\"o}rr, Martin and Potvliege, RM and Shakeshaft, Robin},
  journal={Journal of the Optical Society of America B},
  volume={7},
  number={4},
  pages={433--448},
  year={1990},
  publisher={Optical Society of America}
}

@article{buchleitner1995microwave,
  title={Microwave ionization of three-dimensional hydrogen atoms in a realistic numerical experiment},
  author={Buchleitner, Andreas and Delande, Dominique and Gay, Jean-Claude},
  journal={Journal of the Optical Society of America B},
  volume={12},
  number={4},
  pages={505--519},
  year={1995},
  publisher={Optical Society of America}
}

@article{li2005time,
  title={A time-dependent Hartree--Fock approach for studying the electronic optical response of molecules in intense fields},
  author={Li, Xiaosong and Smith, Stanley M and Markevitch, Alexei N and Romanov, Dmitri A and Levis, Robert J and Schlegel, H Bernhard},
  journal={Physical Chemistry Chemical Physics},
  volume={7},
  number={2},
  pages={233--239},
  year={2005},
  publisher={Royal Society of Chemistry}
}

@article{puchalski2016theoretical,
  title={Theoretical determination of the polarizability dispersion and the refractive index of helium},
  author={Puchalski, Mariusz and Piszczatowski, Konrad and Komasa, Jacek and Jeziorski, Bogumi{\l} and Szalewicz, Krzysztof},
  journal={Physical Review A},
  volume={93},
  number={3},
  pages={032515},
  year={2016},
  publisher={APS}
}

@article{komasa2001dipole,
  title={Dipole and quadrupole polarizabilities and shielding factors of beryllium from exponentially correlated Gaussian functions},
  author={Komasa, Jacek},
  journal={Physical Review A},
  volume={65},
  number={1},
  pages={012506},
  year={2001},
  publisher={APS}
}

@article{runge1984tddft,
  title={Density-Functional Theory for Time-Dependent Systems},
  author={Runge, Erich and Gross, E. K. U.},
  journal={Physical Review Letters},
  volume={52},
  number={12},
  pages={997--1000},
  year={1984},
  doi={10.1103/PhysRevLett.52.997}
}

@article{kreibich2001multicomponent,
  title={Multicomponent Density-Functional Theory for Electrons and Nuclei},
  author={Kreibich, T. and Gross, E. K. U.},
  journal={Physical Review Letters},
  volume={86},
  number={14},
  pages={2984--2987},
  year={2001},
  doi={10.1103/PhysRevLett.86.2984}
}

@article{krausz2009attosecond,
  title={Attosecond physics},
  author={Krausz, Ferenc and Ivanov, Misha},
  journal={Reviews of modern physics},
  volume={81},
  number={1},
  pages={163--234},
  year={2009},
  publisher={APS}
}

@article{corkum2007attosecond,
  title={Attosecond science},
  author={Corkum, Paul B and Krausz, Ferenc},
  journal={Nature physics},
  volume={3},
  number={6},
  pages={381--387},
  year={2007},
  publisher={Nature Publishing Group UK London}
}

@article{zewail2000femtochemistry,
  title={Femtochemistry: atomic-scale dynamics of the chemical bond using ultrafast lasers (Nobel Lecture)},
  author={Zewail, Ahmed H},
  journal={Angewandte Chemie International Edition},
  volume={39},
  number={15},
  pages={2586--2631},
  year={2000},
  publisher={Wiley Online Library}
}

@article{cederbaum1999ultrafast,
  title={Ultrafast charge migration by electron correlation},
  author={Cederbaum, Lorenz S and Zobeley, J},
  journal={Chemical Physics Letters},
  volume={307},
  number={3-4},
  pages={205--210},
  year={1999},
  publisher={Elsevier}
}

@article{nisoli2017attosecond,
  title={Attosecond electron dynamics in molecules},
  author={Nisoli, Mauro and Decleva, Piero and Calegari, Francesca and Palacios, Alicia and Mart{\'\i}n, Fernando},
  journal={Chemical reviews},
  volume={117},
  number={16},
  pages={10760--10825},
  year={2017},
  publisher={ACS Publications}
}

@article{brabec2000intense,
  title={Intense few-cycle laser fields: Frontiers of nonlinear optics},
  author={Brabec, Thomas and Krausz, Ferenc},
  journal={Reviews of Modern Physics},
  volume={72},
  number={2},
  pages={545},
  year={2000},
  publisher={APS}
}

@article{pfau2020,
  title = {Ab Initio Solution of the Many-Electron {Schr\"odinger} Equation with Deep Neural Networks},
  author = {Pfau, David and Foulkes, W. M. C.},
  year = {2020},
  month = sep,
  journal = {Physical Review Research},
  volume = {2},
  number = {3},
  pages = {033429},
  publisher = {American Physical Society},
  doi = {10.1103/PhysRevResearch.2.033429},
  urldate = {2021-01-11}
}

@article{hermann2020deep,
  title={Deep-neural-network solution of the electronic Schr{\"o}dinger equation},
  author={Hermann, Jan and Sch{\"a}tzle, Zeno and No{\'e}, Frank},
  journal={Nature Chemistry},
  volume={12},
  number={10},
  pages={891--897},
  year={2020},
  publisher={Nature Publishing Group UK London}
}

@article{hermann2023ab,
  title={Ab initio quantum chemistry with neural-network wavefunctions},
  author={Hermann, Jan and Spencer, James and Choo, Kenny and Mezzacapo, Antonio and Foulkes, W Matthew C and Pfau, David and Carleo, Giuseppe and No{\'e}, Frank},
  journal={Nature Reviews Chemistry},
  volume={7},
  number={10},
  pages={692--709},
  year={2023},
  publisher={Nature Publishing Group UK London}
}

@article{tang2025deep,
  title={Deep-learning electronic structure calculations},
  author={Tang, Zechen and Chen, Haoxiang and Li, Yang and Qian, Yubing and Wang, Yuxiang and Fu, Weizhong and Li, Jialin and Si, Chen and Duan, Wenhui and Chen, Ji and others},
  journal={Nature Computational Science},
  pages={1--14},
  year={2025},
  publisher={Nature Publishing Group}
}

@article{li2025spin,
  title={Spin-Adapted Neural Network Wavefunctions in Real Space},
  author={Li, Ruichen and Liu, Yuzhi and Jiang, Du and Chen, Yixiao and Wen, Xuelan and Li, Wenrui and He, Di and Wang, Liwei and Chen, Ji and Ren, Weiluo},
  journal={arXiv preprint arXiv:2511.01671},
  year={2025}
}

@article{jiang2025neural,
  title={Neural scaling laws surpass chemical accuracy for the many-electron schr$\backslash$" odinger equation},
  author={Jiang, Du and Wen, Xuelan and Chen, Yixiao and Li, Ruichen and Fu, Weizhong and Pham, Hung Q and Chen, Ji and He, Di and Goddard III, William A and Wang, Liwei and others},
  journal={arXiv preprint arXiv:2508.02570},
  year={2025}
}

@article{fu2025local,
  title={Local pseudopotential unlocks the true potential of neural network-based quantum Monte Carlo},
  author={Fu, Weizhong and Fujimaru, Ryunosuke and Li, Ruichen and Liu, Yuzhi and Wen, Xuelan and Li, Xiang and Hongo, Kenta and Wang, Liwei and Ichibha, Tom and Maezono, Ryo and others},
  journal={arXiv preprint arXiv:2505.19909},
  year={2025}
}

@article{pfau2024excitedstates,
  title={Accurate computation of quantum excited states with neural networks},
  author={Pfau, David and Axelrod, Simon and Sutterud, Halvard and von Glehn, Ingrid and Spencer, James S.},
  journal={Science},
  volume={385},
  number={6711},
  year={2024},
  doi={10.1126/science.adn0137}
}

@article{entwistle2023deepvmc_ex,
  title={Electronic excited states in deep variational Monte Carlo},
  author={Entwistle, M. T. and Sch{\"a}tzle, Z. and Erdman, P. A. and Hermann, J. and No{\'e}, F.},
  journal={Nature Communications},
  volume={14},
  number={1},
  year={2023},
  doi={10.1038/s41467-022-35534-5}
}

@article{orr1971perturbation,
  title={Perturbation theory of the non-linear optical polarization of an isolated system},
  author={Orr, BJ and Ward, JF},
  journal={Molecular Physics},
  volume={20},
  number={3},
  pages={513--526},
  year={1971},
  publisher={Taylor \& Francis}
}

@article{dalgarno1955longrange,
  title={The exact calculation of long-range forces between atoms by perturbation theory},
  author={Dalgarno, Alexander and Lewis, J. T.},
  journal={Proceedings of the Royal Society of London. Series A. Mathematical and Physical Sciences},
  volume={233},
  number={1192},
  pages={70--74},
  year={1955},
  doi={10.1098/rspa.1955.0246}
}

@article{tao2012tsurff,
  title={Photo-electron momentum spectra from minimal volumes: the time-dependent surface flux method},
  author={Tao, Liang and Scrinzi, Armin},
  journal={New Journal of Physics},
  volume={14},
  number={1},
  pages={013021},
  year={2012},
  doi={10.1088/1367-2630/14/1/013021}
}

@article{labeye2018optimal,
  title={Optimal Basis Set for Electron Dynamics in Strong Laser Fields: The case of Molecular Ion H$_2^+$},
  author={Labeye, Marie and Zapata, Felipe and Coccia, Emanuele and V{\'e}niard, Val{\'e}rie and Toulouse, Julien and Caillat, J{\'e}r{\'e}mie and Ta{\"i}eb, Richard and Luppi, Eleonora},
  journal={Journal of Chemical Theory and Computation},
  volume={14},
  number={11},
  pages={5846--5858},
  year={2018},
  doi={10.1021/acs.jctc.8b00656}
}

@article{wozniak2021systematic,
  title={A systematic construction of Gaussian basis sets for the description of laser field ionization and high-harmonic generation},
  author={Wo{\'z}niak, Aleksander P. and Lesiuk, Micha{\l} and Przybytek, Micha{\l} and Efimov, Dmitry K. and Prauzner-Bechcicki, Jakub S. and Mandrysz, Micha{\l} and Ciappina, Marcelo and Pisanty, Emilio and Zakrzewski, Jakub and Lewenstein, Maciej and Moszy{\'n}ski, Robert},
  journal={The Journal of Chemical Physics},
  volume={154},
  number={9},
  pages={094111},
  year={2021},
  doi={10.1063/5.0040879}
}

@article{kvaal2012tdcc,
  title={\textit{Ab initio} quantum dynamics using coupled-cluster},
  author={Kvaal, Simen},
  journal={The Journal of Chemical Physics},
  volume={136},
  number={19},
  pages={194109},
  year={2012},
  doi={10.1063/1.4718427}
}

@article{dreuw2004failure,
  title={Failure of time-dependent density functional theory for long-range charge-transfer excited states: the zincbacteriochlorin- bacteriochlorin and bacteriochlorophyll- spheroidene complexes},
  author={Dreuw, Andreas and Head-Gordon, Martin},
  journal={Journal of the American Chemical Society},
  volume={126},
  number={12},
  pages={4007--4016},
  year={2004},
  publisher={ACS Publications}
}

@article{maitra2004double,
  title={Double excitations within time-dependent density functional theory linear response},
  author={Maitra, Neepa T and Zhang, Fan and Cave, Robert J and Burke, Kieron},
  journal={The Journal of Chemical Physics},
  volume={120},
  number={13},
  pages={5932--5937},
  year={2004},
  publisher={American Institute of Physics}
}

@article{fuks2013dynamics,
  title={Dynamics of charge-transfer processes with time-dependent density functional theory},
  author={Fuks, Johanna I and Elliott, Peter and Rubio, Angel and Maitra, Neepa T},
  journal={The journal of physical chemistry letters},
  volume={4},
  number={5},
  pages={735--739},
  year={2013},
  publisher={ACS Publications}
}

@article{castro2006octopus,
  title={\textit{octopus}: a tool for the application of time-dependent density functional theory},
  author={Castro, Alberto and Appel, Heiko and Oliveira, Micael and Rozzi, Carlo A. and Andrade, Xavier and Lorenzen, Florian and Marques, M. A. L. and Gross, E. K. U. and Rubio, Angel},
  journal={physica status solidi (b)},
  volume={243},
  number={11},
  pages={2465--2488},
  year={2006},
  doi={10.1002/pssb.200642067}
}

@article{andrade2015octopus,
  title={Real-space grids and the Octopus code as tools for the development of new simulation approaches for electronic systems},
  author={Andrade, Xavier and Strubbe, David and De Giovannini, Umberto and Larsen, Ask Hjorth and Oliveira, Micael J. T. and Alberdi-Rodriguez, Joseba and Varas, Alejandro and Theophilou, Iris and Helbig, Nicole and Verstraete, Matthieu J. and Stella, Lorenzo and Nogueira, Fernando and Aspuru-Guzik, Al{\'a}n and Castro, Alberto and Marques, Miguel A. L. and Rubio, Angel},
  journal={Physical Chemistry Chemical Physics},
  volume={17},
  number={47},
  pages={31371--31396},
  year={2015},
  doi={10.1039/C5CP00351B}
}

@article{pavosevic2020neo,
  title={Multicomponent Quantum Chemistry: Integrating Electronic and Nuclear Quantum Effects via the Nuclear--Electronic Orbital Method},
  author={Pavo{\v{s}}evi{\'c}, Fabijan and Culpitt, Tanner and Hammes-Schiffer, Sharon},
  journal={Chemical Reviews},
  volume={120},
  number={9},
  pages={4222--4253},
  year={2020},
  doi={10.1021/acs.chemrev.9b00798}
}

@book{cohentannoudji1992atomphoton,
  title={Atom-Photon Interactions: Basic Processes and Applications},
  author={Cohen-Tannoudji, Claude and Dupont-Roc, Jacques and Grynberg, Gilbert},
  publisher={Wiley-VCH},
  year={1992}
}

@book{szabo1996qc,
  title={Modern Quantum Chemistry: Introduction to Advanced Electronic Structure Theory},
  author={Szabo, Attila and Ostlund, Neil S.},
  publisher={Dover Publications},
  year={1996}
}

@book{ullrich2012tddft,
  title={Time-Dependent Density-Functional Theory: Concepts and Applications},
  author={Ullrich, Carsten A.},
  publisher={Oxford University Press},
  year={2012}
}

@book{landau1977qm,
  title={Quantum Mechanics: Non-Relativistic Theory},
  author={Landau, L. D. and Lifshitz, E. M.},
  publisher={Pergamon Press},
  year={1977},
  edition={3}
}

@article{schrodinger1926wave,
  title={An undulatory theory of the mechanics of atoms and molecules},
  author={Schr{\"o}dinger, Erwin},
  journal={Physical Review},
  volume={28},
  number={6},
  pages={1049--1070},
  year={1926},
  doi={10.1103/PhysRev.28.1049}
}

@article{cai2026permutation,
  title={Permutation invariant multi-scale full quantum neural network wavefunction},
  author={Cai, Pengzhen and Qian, Yubing and Deng, Li and Fu, Weizhong and Yang, Lei and Sun, Zhiyu and Li, Xin-Zheng and Wang, En-Ge and Chen, Liangwen and Ren, Weiluo and others},
  journal={arXiv preprint arXiv:2603.12233},
  year={2026}
}

@article{carr2026neural,
  title={Neural Wavefunction Calculations of $\{$$\backslash$mu$\}$ SR Spectra with Quantum Muons and Protons},
  author={Carr, Jamie and Volkai, Mathias and Foulkes, WMC and Fadon, Andres Perez},
  journal={arXiv preprint arXiv:2603.05453},
  year={2026}
}

@article{carleo2017solving,
  title={Solving the quantum many-body problem with artificial neural networks},
  author={Carleo, Giuseppe and Troyer, Matthias},
  journal={Science},
  volume={355},
  number={6325},
  pages={602--606},
  year={2017},
  publisher={American Association for the Advancement of Science}
}

@article{ding2013efficient,
  title={An efficient method for calculating dynamical hyperpolarizabilities using real-time time-dependent density functional theory},
  author={Ding, Feizhi and Van Kuiken, Benjamin E and Eichinger, Bruce E and Li, Xiaosong},
  journal={The Journal of chemical physics},
  volume={138},
  number={6},
  year={2013},
  publisher={AIP Publishing}
}

@article{fu2024variance,
  title={Variance extrapolation method for neural-network variational Monte Carlo},
  author={Fu, Weizhong and Ren, Weiluo and Chen, Ji},
  journal={Machine Learning: Science and Technology},
  volume={5},
  number={1},
  pages={015016},
  year={2024},
  publisher={IOP Publishing}
}

@article{schutt2017schnet,
  title={Schnet: A continuous-filter convolutional neural network for modeling quantum interactions},
  author={Sch{\"u}tt, Kristof and Kindermans, Pieter-Jan and Sauceda Felix, Huziel Enoc and Chmiela, Stefan and Tkatchenko, Alexandre and M{\"u}ller, Klaus-Robert},
  journal={Advances in neural information processing systems},
  volume={30},
  year={2017}
}

@article{spencer2020better,
  title={Better, faster fermionic neural networks},
  author={Spencer, James S and Pfau, David and Botev, Aleksandar},
  journal={arXiv preprint arXiv:2011.07125},
  year={2020}
}

@article{scherbela2022solving,
  title={Solving the electronic Schr{\"o}dinger equation for multiple nuclear geometries with weight-sharing deep neural networks},
  author={Scherbela, Michael and Reisenhofer, Rafael and Gerard, Leon and Marquetand, Philipp and Grohs, Philipp},
  journal={Nature Computational Science},
  volume={2},
  number={5},
  pages={331--341},
  year={2022},
  publisher={Nature Publishing Group US New York}
}

@article{gao2017efficient,
  title={Efficient representation of quantum many-body states with deep neural networks},
  author={Gao, Xun and Duan, Lu-Ming},
  journal={Nature communications},
  volume={8},
  number={1},
  pages={662},
  year={2017},
  publisher={Nature Publishing Group UK London}
}

@article{cai2018approximating,
  title={Approximating quantum many-body wave functions using artificial neural networks},
  author={Cai, Zi and Liu, Jinguo},
  journal={Physical Review B},
  volume={97},
  number={3},
  pages={035116},
  year={2018},
  publisher={APS}
}

@article{han2019solving,
  title={Solving many-electron Schr{\"o}dinger equation using deep neural networks},
  author={Han, Jiequn and Zhang, Linfeng and Weinan, E},
  journal={Journal of Computational Physics},
  volume={399},
  pages={108929},
  year={2019},
  publisher={Elsevier}
}

@article{luo2019backflow,
  title={Backflow transformations via neural networks for quantum many-body wave functions},
  author={Luo, Di and Clark, Bryan K},
  journal={Physical review letters},
  volume={122},
  number={22},
  pages={226401},
  year={2019},
  publisher={APS}
}

@article{von2022self,
  title={A self-attention ansatz for ab-initio quantum chemistry},
  author={von Glehn, Ingrid and Spencer, James S and Pfau, David},
  journal={arXiv preprint arXiv:2211.13672},
  year={2022}
}

@article{lin2023explicitly,
  title={Explicitly antisymmetrized neural network layers for variational Monte Carlo simulation},
  author={Lin, Jeffmin and Goldshlager, Gil and Lin, Lin},
  journal={Journal of Computational Physics},
  volume={474},
  pages={111765},
  year={2023},
  publisher={Elsevier}
}

@article{abrahamsen2022taming,
  title={Taming the sign problem of explicitly antisymmetrized neural networks via rough activation functions},
  author={Abrahamsen, Nilin and Lin, Lin},
  journal={arXiv preprint arXiv:2205.12250},
  year={2022}
}

@article{gerard2022gold,
  title={Gold-standard solutions to the Schr{\"o}dinger equation using deep learning: How much physics do we need?},
  author={Gerard, Leon and Scherbela, Michael and Marquetand, Philipp and Grohs, Philipp},
  journal={Advances in Neural Information Processing Systems},
  volume={35},
  pages={10282--10294},
  year={2022}
}

@article{pescia2022neural,
  title={Neural-network quantum states for periodic systems in continuous space},
  author={Pescia, Gabriel and Han, Jiequn and Lovato, Alessandro and Lu, Jianfeng and Carleo, Giuseppe},
  journal={Physical Review Research},
  volume={4},
  number={2},
  pages={023138},
  year={2022},
  publisher={APS}
}

@article{wilson2023neural,
  title={Neural network ansatz for periodic wave functions and the homogeneous electron gas},
  author={Wilson, Max and Moroni, Saverio and Holzmann, Markus and Gao, Nicholas and Wudarski, Filip and Vegge, Tejs and Bhowmik, Arghya},
  journal={Physical Review B},
  volume={107},
  number={23},
  pages={235139},
  year={2023},
  publisher={APS}
}

@article{gao2021ab,
  title={Ab-initio potential energy surfaces by pairing GNNs with neural wave functions},
  author={Gao, Nicholas and G{\"u}nnemann, Stephan},
  journal={arXiv preprint arXiv:2110.05064},
  year={2021}
}

@article{ren2023towards,
  title={Towards the ground state of molecules via diffusion Monte Carlo on neural networks},
  author={Ren, Weiluo and Fu, Weizhong and Wu, Xiaojie and Chen, Ji},
  journal={Nature Communications},
  volume={14},
  number={1},
  pages={1860},
  year={2023},
  publisher={Nature Publishing Group UK London}
}

@article{li2022,
  title = {Ab Initio Calculation of Real Solids via Neural Network Ansatz},
  author = {Li, Xiang and Li, Zhe and Chen, Ji},
  year = {2022},
  month = dec,
  journal = {Nature Communications},
  volume = {13},
  number = {1},
  pages = {7895},
  publisher = {Nature Publishing Group},
  issn = {2041-1723},
  doi = {10.1038/s41467-022-35627-1},
  urldate = {2023-02-14},
  copyright = {2022 The Author(s)},
  langid = {english}
}

@article{gao2022sampling,
  title={Sampling-free inference for ab-initio potential energy surface networks},
  author={Gao, Nicholas and G{\"u}nnemann, Stephan},
  journal={arXiv preprint arXiv:2205.14962},
  year={2022}
}

@article{gao2023generalizing,
  title={Generalizing neural wave functions},
  author={Gao, Nicholas and G{\"u}nnemann, Stephan},
  journal={arXiv preprint arXiv:2302.04168},
  year={2023}
}

@article{scherbela2023towards,
  title={Towards a Foundation Model for Neural Network Wavefunctions},
  author={Scherbela, Michael and Gerard, Leon and Grohs, Philipp},
  journal={arXiv preprint arXiv:2303.09949},
  year={2023}
}

@article{austin2012quantum,
  title={Quantum Monte Carlo and related approaches},
  author={Austin, Brian M and Zubarev, Dmitry Yu and Lester Jr, William A},
  journal={Chemical reviews},
  volume={112},
  number={1},
  pages={263--288},
  year={2012},
  publisher={ACS Publications}
}

@article{kashima2001path,
  title={Path-integral renormalization group method for numerical study on ground states of strongly correlated electronic systems},
  author={Kashima, Tsuyoshi and Imada, Masatoshi},
  journal={journal of the physical society of japan},
  volume={70},
  number={8},
  pages={2287--2299},
  year={2001},
  publisher={The Physical Society of Japan}
}

@article{choo2020fermionic,
  title={Fermionic neural-network states for ab-initio electronic structure},
  author={Choo, Kenny and Mezzacapo, Antonio and Carleo, Giuseppe},
  journal={Nature communications},
  volume={11},
  number={1},
  pages={2368},
  year={2020},
  publisher={Nature Publishing Group UK London}
}

@article{szabo2024improved,
  title={An improved penalty-based excited-state variational Monte Carlo approach with deep-learning ansatzes},
  author={Szab{\'o}, P Bern{\'a}t and Scha{\"a}tzle, Zeno and Entwistle, Michael T and No{\'e}, Frank},
  journal={Journal of Chemical Theory and Computation},
  volume={20},
  number={18},
  pages={7922--7935},
  year={2024},
  publisher={ACS Publications}
}

@article{schatzle2025ab,
  title={Ab-initio simulation of excited-state potential energy surfaces with transferable deep quantum Monte Carlo},
  author={Sch{\"a}tzle, Zeno and Szab{\'o}, P Bern{\'a}t and Cuzzocrea, Alice and No{\'e}, Frank},
  journal={arXiv preprint arXiv:2503.19847},
  year={2025}
}

@article{gao2026excited,
  title={Excited Pfaffians: Generalized Neural Wave Functions Across Structure and State},
  author={Gao, Nicholas and Grutschus, Till and No{\'e}, Frank and G{\"u}nnemann, Stephan},
  journal={arXiv preprint arXiv:2603.14515},
  year={2026}
}

@article{li2024spin,
  title={Spin-symmetry-enforced solution of the many-body Schr{\"o}dinger equation with a deep neural network},
  author={Li, Zhe and Lu, Zixiang and Li, Ruichen and Wen, Xuelan and Li, Xiang and Wang, Liwei and Chen, Ji and Ren, Weiluo},
  journal={Nature Computational Science},
  volume={4},
  number={12},
  pages={910--919},
  year={2024},
  publisher={Nature Publishing Group US New York}
}

@article{li2024computational,
  title={A computational framework for neural network-based variational Monte Carlo with Forward Laplacian},
  author={Li, Ruichen and Ye, Haotian and Jiang, Du and Wen, Xuelan and Wang, Chuwei and Li, Zhe and Li, Xiang and He, Di and Chen, Ji and Ren, Weiluo and others},
  journal={Nature Machine Intelligence},
  volume={6},
  number={2},
  pages={209--219},
  year={2024},
  publisher={Nature Publishing Group UK London}
}

@article{muga2004complex,
  title={Complex absorbing potentials},
  author={Muga, J Gonzalo and Palao, JP and Navarro, B and Egusquiza, IL},
  journal={Physics Reports},
  volume={395},
  number={6},
  pages={357--426},
  year={2004},
  publisher={Elsevier}
}

@article{barrett2022autoregressive,
  title={Autoregressive neural-network wavefunctions for ab initio quantum chemistry},
  author={Barrett, Thomas D and Malyshev, Aleksei and Lvovsky, AI},
  journal={Nature Machine Intelligence},
  volume={4},
  number={4},
  pages={351--358},
  year={2022},
  publisher={Nature Publishing Group UK London}
}

@article{zhao2023scalable,
  title={Scalable neural quantum states architecture for quantum chemistry},
  author={Zhao, Tianchen and Stokes, James and Veerapaneni, Shravan},
  journal={Machine Learning: Science and Technology},
  year={2023}
}

@article{hackl2020geometry,
  title={Geometry of variational methods: dynamics of closed quantum systems},
  author={Hackl, Lucas and Guaita, Tommaso and Shi, Tao and Haegeman, Jutho and Demler, Eugene and Cirac, J Ignacio},
  journal={SciPost Physics},
  volume={9},
  number={4},
  pages={048},
  year={2020}
}

@article{orr1971nonlinear,
  title={Perturbation theory of the non-linear optical polarization of an isolated system},
  author={Orr, B. J. and Ward, J. F.},
  journal={Molecular Physics},
  volume={20},
  number={3},
  pages={513--526},
  year={1971},
  doi={10.1080/00268977100100481}
}

@article{sun2018pyscf,
  title={PySCF: the Python-based simulations of chemistry framework},
  author={Sun, Qiming and Berkelbach, Timothy C and Blunt, Nick S and Booth, George H and Guo, Sheng and Li, Zhendong and Liu, Junzi and McClain, James D and Sayfutyarova, Elvira R and Sharma, Sandeep and others},
  journal={Wiley Interdisciplinary Reviews: Computational Molecular Science},
  volume={8},
  number={1},
  pages={e1340},
  year={2018},
  publisher={Wiley Online Library}
}

@article{sun2020recent,
  title={Recent developments in the PySCF program package},
  author={Sun, Qiming and Zhang, Xing and Banerjee, Samragni and Bao, Peng and Barbry, Marc and Blunt, Nick S and Bogdanov, Nikolay A and Booth, George H and Chen, Jia and Cui, Zhi-Hao and others},
  journal={The Journal of chemical physics},
  volume={153},
  number={2},
  year={2020},
  publisher={AIP Publishing}
}

@article{krause2007molecular,
  title={Molecular response properties from explicitly time-dependent configuration interaction methods},
  author={Krause, Pascal and Klamroth, Tillmann and Saalfrank, Peter},
  journal={The Journal of chemical physics},
  volume={127},
  number={3},
  year={2007},
  publisher={AIP Publishing}
}

@article{broeckhove1988equivalence,
  title={On the equivalence of time-dependent variational principles},
  author={Broeckhove, J and Lathouwers, L and Kesteloot, E and Van Leuven, P},
  journal={Chemical physics letters},
  volume={149},
  number={5-6},
  pages={547--550},
  year={1988},
  publisher={Elsevier}
}

@article{yuan2019theory,
  title={Theory of variational quantum simulation},
  author={Yuan, Xiao and Endo, Suguru and Zhao, Qi and Li, Ying and Benjamin, Simon C},
  journal={Quantum},
  volume={3},
  pages={191},
  year={2019},
  publisher={Verein zur F{\"o}rderung des Open Access Publizierens in den Quantenwissenschaften}
}

\end{document}


\title{Supplementary Information: Towards stable and accurate electron dynamics via neural network based time-dependent variational Monte Carlo}

\author{ }

\date{\vspace{-5ex}}
\maketitle
\tableofcontents
\newpage

\section{Hyperparameters}
The detailed hyperparameters of the neural network training and evolving are summarized in \Cref{tab:hyperparams}.
Note that the specified number of determinants refers to the total count of the network rather than the number per state, that is, we employ only a single determinant to represent each individual state during the NES-VMC training.
The number of Markov chain Monte Carlo (MCMC) steps per iteration is set to 30 for both the training and evolution processes.
Finally, to minimize statistical variance and mitigate the accumulation of errors during time evolution, the batch size is chosen to be much larger than that used for training.
Notably, the batch size for the NB-tVMC calculation in H$_2$ system is set to as large as 327680 for a long time stable evolution under the relatively strong electric field.

\begin{table}[htbp]
\centering
\caption{\label{tab:hyperparams} Default hyperparameters of neural network training and evolving.}
\begin{tabular}{ccc}
\hline \hline
Source & Parameter & Value \\
\hline
\multirow{4}{*}{Network Architecture}
  & Determinants                    & 5  \\
  & Network layers                  & 4 \\
  & One-electron stream width                & 64 \\
  & Two-electron stream width                & 16 \\
\hline
MCMC & Steps & 30 \\
\hline
\multirow{4}{*}{Training}
  & Batch size & 20480 \\
  & Iterations & 50000 \\
  & Number of states & 5 \\
  & Spin penalty strength (a.u.) & 1.0 \\
\hline
\multirow{2}{*}{Evolving}
  & Batch size & 81920 \\
  & Time step (a.u.) & 0.001 \\
\hline \hline
\end{tabular}
\end{table}

\section{Methodological Details}
The NB-tVMC time evolution is obtained from the Dirac-Frenkel's principle (DFP), which projects the Schr\"{o}dinger equation onto the tangent space of the variational ansatz.
Let $\boldsymbol{\theta}(t)$ denote the (generally complex) variational parameters of the wavefunction $\Psi(\mathbf{x};\boldsymbol{\theta})$.
Defining the logarithmic derivatives
\begin{equation}
  O_i(\mathbf{x}) \equiv \partial_{\theta_i}\ln \Psi(\mathbf{x};\boldsymbol{\theta}),
\end{equation}
and the local energy
\begin{equation}
  E_{\mathrm{L}}(\mathbf{x}) \equiv \frac{\hat H\Psi(\mathbf{x};\boldsymbol{\theta})}{\Psi(\mathbf{x};\boldsymbol{\theta})},
\end{equation}
the equation of motion can be written as a linear system
\begin{equation}
  \sum_j S_{ij}\,\dot{\theta}_j = -i\,F_i.
  \label{eq:tdvp_linear}
\end{equation}
Here $\mathbf{S}$ is the (stochastic) quantum geometric tensor (QGT) and $\mathbf{F}$ is the ``force'' vector. In tVMC they are evaluated stochastically by Monte Carlo sampling,
\begin{align}
  S_{ij} &= \big\langle O_i^* O_j\big\rangle - \big\langle O_i^*\big\rangle\big\langle O_j\big\rangle, \\
  F_i &= \big\langle O_i^* E_{\mathrm{L}}\big\rangle - \big\langle O_i^*\big\rangle\big\langle E_{\mathrm{L}}\big\rangle,
\end{align}
where $\langle\cdots\rangle$ denotes an expectation value over $|\Psi(\mathbf{x};\boldsymbol{\theta})|^2$.

\begin{figure}[htbp]
  \centering
  \includegraphics[width=0.5\textwidth]{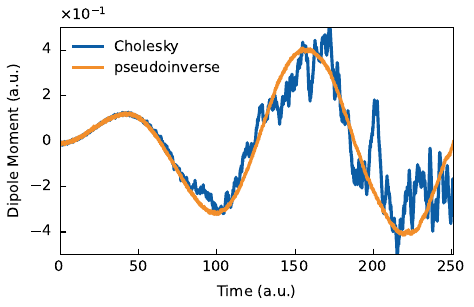}
  \caption{Comparison of solving \Cref{eq:tdvp_linear} using the Moore--Penrose pseudoinverse and Cholesky decomposition methods.}
  \label{fig:inverse}
\end{figure}

In practice, $\mathbf{S}$ can be ill-conditioned due to finite sampling noise and redundant directions in parameter space.
We thus solve \Cref{eq:tdvp_linear} using the Moore--Penrose pseudoinverse of $\mathbf{S}$, computed via singular value decomposition (SVD) with a cutoff on small singular values.
We also attempted the computationally faster Cholesky decomposition solver, but it exhibited substancially poorer numerical stability, as illustrated in \Cref{fig:inverse}.
Time integration of $\dot{\boldsymbol{\theta}}$ is performed using a second-order Heun (predictor--corrector) scheme:
\begin{align}
  \boldsymbol{\theta}^{(p)} &= \boldsymbol{\theta}^{(n)} + \Delta t\,\dot{\boldsymbol{\theta}}^{(n)}, \\
  \boldsymbol{\theta}^{(n+1)} &= \boldsymbol{\theta}^{(n)} + \tfrac{\Delta t}{2}\Big(\dot{\boldsymbol{\theta}}^{(n)} + \dot{\boldsymbol{\theta}}^{(p)}\Big),
\end{align}
where $\dot{\boldsymbol{\theta}}^{(p)}$ is obtained by re-evaluating $\mathbf{S}$ and $\mathbf{F}$ at $\boldsymbol{\theta}^{(p)}$.

All calculations are performed in double precision, and sufficiently large Monte Carlo batch sizes are used during propagation to reduce variance and its accumulation over time.
During the NES-VMC multistate training stage, a spin-penalty term is used to bias the optimization toward the desired spin sector.
Finally, to obtain a higher-quality ground state within the fixed manifold, we follow the NES-VMC-inferred linear mixing coefficients by an additional imaginary-time projection that updates only the final linear layer coefficients.

We further give a discussion about neural basis manifold. In NB-tVMC, the nonlinear part of the neural network (which defines the neural basis) is frozen during time propagation, and only the final linear layer is propagated.
We adopt the same notation as in the main text, where the neural basis functions $\chi^{\alpha}_{\mu j}(\mathbf{x})$ are outputs of the last hidden layer and only the final linear layer parameters are time dependent.
To make explicit that this restriction corresponds to a \emph{fixed linear subspace} of many-body wavefunctions, define a feature matrix $\mathbf{X}^{\alpha}(\mathbf{x})\in\mathbb{R}^{M\times N_\alpha}$ with entries $X^{\alpha}_{\mu j}=\chi^{\alpha}_{\mu j}$ and a coefficient matrix $\tilde{\boldsymbol{\Theta}}^{k\alpha}(t)\in\mathbb{R}^{N_\alpha\times M}$ collecting the time-dependent linear coefficients.
Then
\begin{equation}
  \boldsymbol{\Phi}^{k\alpha}(t) = \tilde{\boldsymbol{\Theta}}^{k\alpha}(t)\,\mathbf{X}^{\alpha}(\mathbf{x}).
  \label{eq:phi_WH}
\end{equation}
With $\mathbf{X}^{\alpha}$ fixed, the Cauchy--Binet formula gives an explicit configuration-like expansion,
\begin{equation}
  \det\big(\boldsymbol{\Phi}^{k\alpha}(t)\big)
  = \sum_{\substack{I\subset\{1,\ldots,M\}\\ |I|=N_\alpha}}
  \det\big(\tilde{\boldsymbol{\Theta}}^{k\alpha}_{:,I}(t)\big)\,\det\big((\mathbf{X}^{\alpha})_{I,:}\big),
  \label{eq:cauchy_binet}
\end{equation}
where $\tilde{\boldsymbol{\Theta}}^{k\alpha}_{:,I}$ is the $N_\alpha\times N_\alpha$ submatrix formed by columns indexed by $I$ and $(\mathbf{X}^{\alpha})_{I,:}$ is the corresponding $N_\alpha\times N_\alpha$ submatrix.
Equation~\eqref{eq:cauchy_binet} shows that, for each spin channel, propagating only the final linear layer amounts to evolving only the coefficients in front of a \emph{fixed} set of determinant basis functions $\det\big((\mathbf{X}^{\alpha})_{I,:}\big)$.
Taking the product of the spin-up and spin-down expansions and summing over $k$, the full wavefunction can be written as a linear combination of fixed many-body basis functions of the form
$\det\big((\mathbf{X}^{\uparrow})_{I,:}\big)\det\big((\mathbf{X}^{\downarrow})_{J,:}\big)$,
with time-dependent coefficients given by products of minors of $\tilde{\boldsymbol{\Theta}}^{k\uparrow}(t)$ and $\tilde{\boldsymbol{\Theta}}^{k\downarrow}(t)$.
Therefore, NB-tVMC confines the dynamics to a compact \emph{linear subspace} of the full many-body Hilbert space determined by the trained neural basis.

In practical real-space neural wavefunctions, orbitals may be multiplied by envelope functions to enforce correct asymptotic decay.
If the envelope is orbital dependent (i.e., applied elementwise in the orbital index), the strict product form \eqref{eq:phi_WH} is no longer exact, and the decomposition \eqref{eq:cauchy_binet} becomes approximate.
We tested an alternative design that applies the envelope at the feature level to restore an exact product structure; it did not lead to a noticeable improvement for the systems studied here, and we thus adopt the standard choice.

\section{Numerical Instabilities in Full-parameter Time Evolution}
\label{sec:principles}
We now discuss why a direct tVMC propagation of all parameters in an expressive neural network wavefunction is numerically unstable, and why the neural basis restriction used in NB-tVMC avoids this difficulty.
For a time-dependent variational state $\Psi(\mathbf{x};\boldsymbol{\theta}(t))$, several closely related variational principles can be used to project the Schr\"{o}dinger equation onto the tangent space of the ansatz~\cite{hackl2020geometry,broeckhove1988equivalence,yuan2019theory}.
For an unnormalized wavefunction, this projection should be understood in projective Hilbert space.
As emphasized in Sec.~3.2 of Ref.~\cite{hackl2020geometry}, physical states are rays, $|\Psi\rangle\sim c|\Psi\rangle$, and tangent vectors that differ by a component proportional to $|\Psi\rangle$ represent the same physical change.
Therefore, one should choose the orthogonal representative of each tangent vector using the projector
\begin{equation}
  \mathbb{Q}_\Psi |\delta\Psi\rangle
  = |\delta\Psi\rangle
  - |\Psi\rangle\frac{\braket{\Psi|\delta\Psi}}{\braket{\Psi|\Psi}} .
  \label{eq:projective_tangent_projector}
\end{equation}
For the variational tangent vector $\partial_{\theta_i}\Psi=O_i\Psi$, this gives
\begin{equation}
  \mathbb{Q}_\Psi\partial_{\theta_i}\Psi
  =\big(O_i-\langle O_i\rangle\big)\Psi.
  \label{eq:mean_shift_tangent}
\end{equation}
The projected residual entering the variational principles can then be written directly as
\begin{equation}
  \mathcal{R}_i(\dot{\boldsymbol{\theta}})
  \equiv \frac{\braket{\mathbb{Q}_\Psi\partial_i\Psi|\hat{K}\Psi}}{\braket{\Psi|\Psi}}
  = F_i - i\sum_j S_{ij}\dot{\theta}_j,
  \label{eq:projected_residual}
\end{equation}
where $\mathbf{S}$ and $\mathbf{F}$ are stochastic QGT and force vector defined in Supplementary Note 2, and $\hat{K} \equiv \hat{H} - i\partial_t$.
The three principles differ only in which part of this complex residual is required to vanish:
\begin{subequations}
\label{eq:three_principles}
\begin{align}
\text{McLachlan's principle (MP):}\,\,
&\min_{\dot{\boldsymbol{\theta}}}\,\big\|i\partial_t\Psi-\hat{H}\Psi\big\|^2
\,\Longleftrightarrow\,
\mathrm{Im}\,\mathcal{R}_i=0
\,\Longleftrightarrow\,
\sum_j \mathrm{Re}\,S_{ij}\dot{\theta}_j
=\mathrm{Re}[-iF_i],
\label{eq:mclachlan_principle}
\\[1.0em]
\text{Time dependent variational principle (TDVP):}\,\,
&\delta\int dt\,\braket{\Psi|\hat{K}|\Psi}=0
\,\Longleftrightarrow\,
\mathrm{Re}\,\mathcal{R}_i=0
\,\Longleftrightarrow\,
\sum_j \mathrm{Im}\,S_{ij}\dot{\theta}_j
=\mathrm{Im}[-iF_i],
\label{eq:tdvp_principle}
\\[1.0em]
\text{Dirac-Frenkel's principle (DFP):}\,\,
&\mathcal{R}_i=0
\,\Longleftrightarrow\,
\sum_j S_{ij}\dot{\theta}_j=-iF_i .
\label{eq:dfp_principle}
\end{align}
\end{subequations}
Here MP and TDVP are written for real variational parameters, so they impose one real condition on each tangent direction.
DFP imposes the full complex condition and is naturally formulated for complex parameters.
This complexification is well defined only when the ansatz is holomorphic with respect to all variational parameters.
When this condition is satisfied, the three formulations are mathematically equivalent: by decomposing every complex DFP parameter as $\theta_i=\theta_i^{\mathrm{R}}+i\theta_i^{\mathrm{I}}$ and treating $\theta_i^{\mathrm{R}}$ and $\theta_i^{\mathrm{I}}$ as independent real parameters, one can recover the corresponding real-parameter MP/TDVP equations.
For a non-holomorphic parametrization, however, DFP is no longer directly applicable.
One must instead propagate real parameters using either MP or TDVP, and the two projected dynamics are then generally inequivalent.
Among them, TDVP preserves the symplectic structure of the variational dynamics and therefore conserves the energy when the Hamiltonian is time independent, whereas MP does not in general possess this long-time stability property.

\begin{figure}[htbp]
  \centering
  \includegraphics[width=0.9\textwidth]{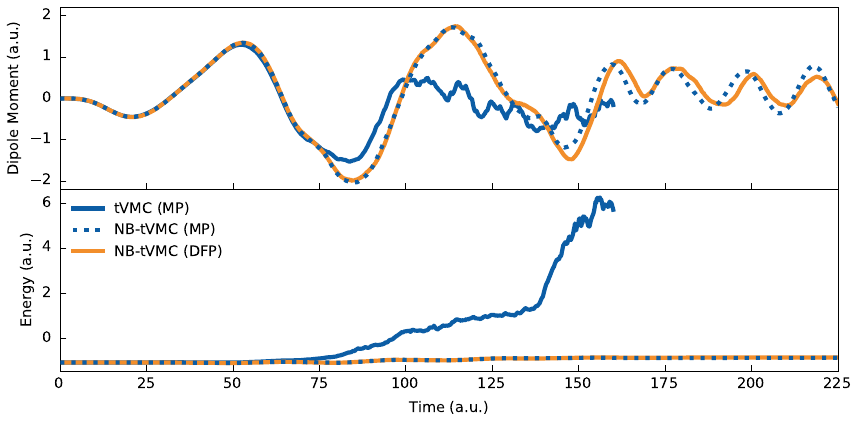}
  \caption{Evolution trajectories of the dipole moment and energy under different parameter-update schemes. The standard tVMC calculations update all neural network parameters using MP, whereas NB-tVMC updates only the final linear layer using MP or DFP.}
  \label{fig:tdvp_si}
\end{figure}
For the FermiNet-type ansatz considered in this work, the dependence on network parameters is holomorphic except at the isolated poles of the $\tanh$ activation function in the complex plane.
We therefore tested the update schemes above directly, as shown in \Cref{fig:tdvp_si}.
The TDVP result is not displayed because it becomes numerically singular already at the first time step: the initial wavefunction is real, making the imaginary part of the QGT nearly a zero matrix and hence impossible to invert in a stable manner.
The full-parameter DFP evolution also breaks down rapidly (within ten time steps) and is not displayed.
The likely origin is that once all neural network parameters are allowed to move in the complex plane, the nonlinear layers can approach regions where the analytically continued $\tanh$ activation is no longer bounded, leading to a rapid amplification of numerical errors.

We examined a full-parameter MP evolution, denoted as tVMC (MP), in which the parameters entering the nonlinear layers are constrained to remain real while the final linear layer coefficients are represented by their real and imaginary parts.
This avoids the most severe complex-plane instability of the nonlinear network and indeed extends the propagation time compared with tVMC (DFP).
Nevertheless, the trajectory eventually becomes unstable.
This behavior is consistent with the fact that MP does not preserve the symplectic structure of the exact Schr\"{o}dinger dynamics and therefore cannot guarantee stable long-time energy behavior, especially when stochastic noise from Monte Carlo sampling is accumulated over many integration steps.

As a cross-check, we finally restrict the time evolution to the neural basis manifold and update only the final linear layer, i.e., the NB-tVMC setting used throughout the main text.
In this case MP and DFP are expected to give equivalent projected dynamics.
The two NB-tVMC curves in \Cref{fig:tdvp_si} are indeed consistent with each other for both the dipole moment and the energy, and both remain stable over the full propagation time.
This comparison supports the central design of NB-tVMC: by freezing the nonlinear neural basis and evolving only a compact set of linear coefficients, the method retains the expressiveness of the trained neural manifold while eliminating the dominant numerical instabilities associated with full-parameter real-time evolution.

\section{Sum-over-states Expression for the Dynamic Polarizability}
Following the definition in the main text, we consider a weak monochromatic field polarized along $z$, $E(t)=E_{\mathrm{max}}\sin(\omega t)$, coupled through $\hat V(t)=-\hat\mu_z E(t)$.
In linear response, $\alpha(\omega)$ is defined as the proportionality constant between the field amplitude $E_{\mathrm{max}}$ and the component of the induced dipole oscillating at the driving frequency $\omega$.
Using time-dependent perturbation theory in the eigenbasis of $\hat H_0|n\rangle=E_n|n\rangle$ (with ground state $|0\rangle$), one obtains the standard sum-over-states representation~\cite{orr1971nonlinear,dalgarno1955longrange}:
\begin{equation}
    \alpha(\omega)
  = 2\sum_{n\ne 0}
  \frac{(E_n-E_0)\,|\langle 0|\hat\mu_z|n\rangle|^2}{(E_n-E_0)^2-\omega^2}
  \, +\, 2\,\mathcal{P}\!\int_{E_\mathrm{ion}}^{\infty}\! dE\,
  \frac{(E-E_0)\,|\langle 0|\hat\mu_z|E\rangle|^2}{(E-E_0)^2-\omega^2}.
  \label{eq:alpha_sos}
\end{equation}
Here $\mathcal{P}$ denotes the Cauchy principal value~\cite{landau1977qm}, which regularizes the integral at the resonant points where the denominator vanishes.
The discrete sum accounts for bound excited states, while the integral captures the continuum spectrum above the ionization threshold $E_\mathrm{ion}$.

To connect \Cref{eq:alpha_sos} to a finite Hilbert-space representation, let $\mathcal{S}$ denote a truncated subspace spanned by a finite set of configuration states $|\tilde{n}\rangle$.
Restricting the linear-response dynamics to $\mathcal{S}$ leads to a truncated polarizability
\begin{equation}
  \alpha_{\mathcal{S}}(\omega)
  = 2\sum_{\tilde{n}\ne \tilde{0}}
  \frac{(\tilde{E}_n-\tilde{E}_0)\,|\langle \tilde{0}|\hat\mu_z|\tilde{n}\rangle|^2}{(\tilde{E}_n-\tilde{E}_0)^2-\omega^2}.
  \label{eq:alpha_sos_truncated}
\end{equation}
In the off-resonant regime relevant for stable fitting (in particular, when $\omega$ lies below the lowest excitation and below the ionization threshold), each contribution is non-negative, which implies
\begin{equation}
  0\le \alpha_{\mathcal{S}}(\omega) \le \alpha(\omega).
  \label{eq:alpha_inequality}
\end{equation}
Therefore, a truncated representation typically tends to underestimate the dynamic polarizability relative to the exact result, which is consistent with the observed underestimation of peak amplitudes in inaccurate evolution trajectories.

\section{Electric Field Shapes}
\begin{figure}[htbp]
  \centering
  \begin{minipage}[b]{0.33\textwidth}
    \centering
    \includegraphics[width=\textwidth]{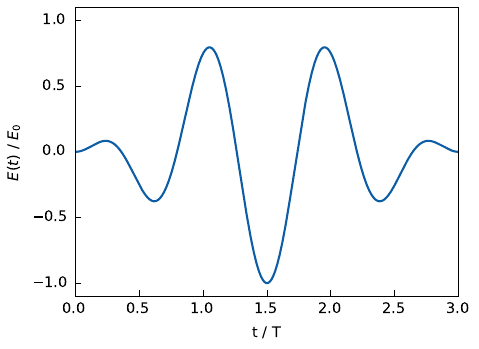}
  \end{minipage}
  \begin{minipage}[b]{0.33\textwidth}
    \centering
    \includegraphics[width=\textwidth]{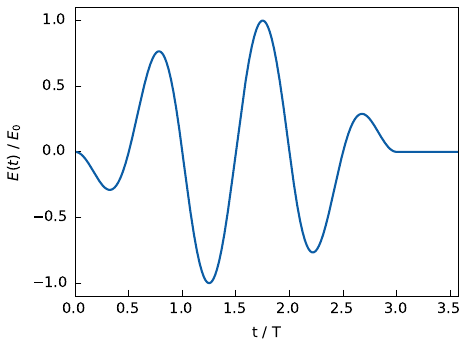}
  \end{minipage}
  \begin{minipage}[b]{0.33\textwidth}
    \centering
    \includegraphics[width=\textwidth]{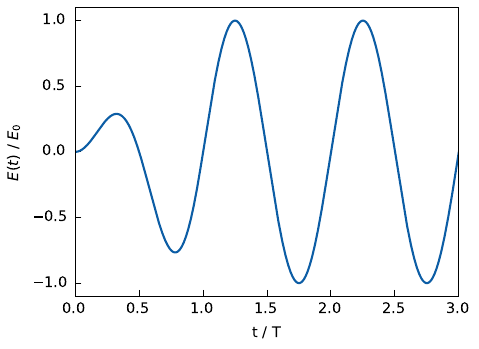}
  \end{minipage}
  \caption{Shapes of the external electric fields employed in the main text for the H atom, H$_2$ molecule, and He/Be atom systems (from left to right).}
  \label{fig:efield_shapes}
\end{figure}

The shapes of the external electric fields employed in the main text for the hydrogen atom, hydrogen molecule, and He/Be atom systems are shown in \Cref{fig:efield_shapes}.
The hydrogen atom system is subjected to an external laser pulse with a sine-squared envelope, defined by
\begin{equation}
  \mathbf{E}(t) = E_{\mathrm{max}} \left[ \sin^2\left(\frac{\pi t}{3T}\right) \cos(\omega t) \right. 
  \left. + \frac{\pi}{3\omega T} \sin\left(\frac{2\pi t}{3T}\right) \sin(\omega t) \right] \hat{\mathbf{z}},
\end{equation}
where $E_{\mathrm{max}} = 0.029$~a.u., $\omega = 0.057$~a.u., and $T = 2\pi/\omega$~\cite{han2010comparison}.
The hydrogen molecule system is driven by a laser pulse with a trapezoidal envelope, defined by
\begin{equation}
\mathbf{E}(t) = -E_{\mathrm{max}} \sin (\omega t) \hat{\mathbf{z}} \times 
\begin{cases} 
t/T & (0 \le t < T) \\
1 & (T \le t < 2T) \\
3 - t/T & (2T \le t < 3T) \\
\end{cases},
\end{equation}
where $E_{\mathrm{max}} = 0.07$~a.u., $\omega = 0.1$~a.u., and $T = 2\pi/\omega$~\cite{li2005time}.
The electric field employed to calculate the dynamic polarizability of He/Be atoms is ramped up with a linear envelope over one optical cycle.

\section{Dipole Moment Results: During Evolution versus Inference}
\begin{figure}[htbp]
  \centering
  \includegraphics[width=\textwidth]{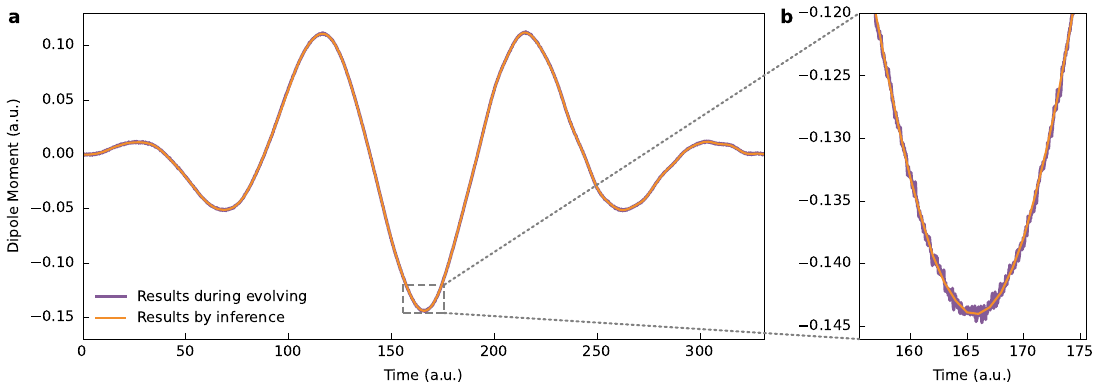}
  \caption{\textbf{(a)} Comparison of dipole moment results during evolving and by inference. \textbf{(b)} Zoom-in view of the highest peak region.}
  \label{fig:inference}
\end{figure}

In \Cref{fig:inference}, we use the H atom system as an example to compare the dipole moment calculated on-the-fly during time evolution with the results obtained via separate inference.
The dipole moment during the evolution process is estimated using samples from the second half of each MCMC sampling round.
Specifically, it is calculated as the average dipole moment of all samples from the last 15 steps of a 30-step MCMC sampling procedure.
To reduce variance, the results presented in the figure are further smoothed using a moving average over a window of 100 time steps.
In contrast, the inference results are obtained by performing 50,000 MCMC sampling steps on the checkpoint wavefunctions, which were saved at 1 a.u. time intervals.
As illustrated in the figure, the two trajectories align with each other perfectly.
This strong agreement demonstrates that the number of MCMC steps chosen during the evolution process is sufficient to ensure a stable and converged sampling distribution.

In the main text, inference results were explicitly calculated to quantitatively evaluate the RMSE in the H atom system.
The results in all other systems were obtained directly during the evolution process, which is entirely sufficient for qualitative analysis.





\section{Details of Reference Real-Time Calculations}
\label{sec:reference_realtime_calculations}

Here we provide the computational details for the deterministic reference calculations used in the main text, including the FCI calculations shown for the hydrogen atom and stretched H$_2$ molecule, and the grid-based RT-TDDFT calculation used as an independent real-space reference for the stretched H$_2$ molecule.

The FCI reference trajectories are obtained by propagating the many-body wavefunction in a finite Gaussian-orbital Hilbert space, following the time-dependent CI formulation of Ref.~\cite{krause2007molecular}.
For each basis set used in the main text, we first perform a conventional FCI calculation with the \textsc{PySCF} package~\cite{sun2018pyscf,sun2020recent} to obtain the field-free FCI eigenstates, their energies, and the transition dipole matrix in this eigenstate basis.
The time-dependent wavefunction is expanded as
\begin{equation}
  |\Psi(t)\rangle = \sum_n C_n(t)|\Phi_n\rangle,
  \label{eq:fci_ci_expansion}
\end{equation}
where $|\Phi_n\rangle$ are the field-free FCI eigenstates.
Under the dipole approximation, the Hamiltonian matrix in this basis reads
\begin{equation}
  H_{mn}(t) = E_m\delta_{mn} - \boldsymbol{\mu}_{mn}\cdot\mathbf{E}(t),
  \label{eq:fci_td_hamiltonian}
\end{equation}
where $E_m$ are the field-free FCI energies and $\boldsymbol{\mu}_{mn}=\langle m|\hat{\boldsymbol{\mu}}|n\rangle$ are the transition dipole matrix elements.
The expansion coefficients $\mathbf{C}(t)$ then satisfy
\begin{equation}
  i\frac{d}{dt}\mathbf{C}(t)=\mathbf{H}(t)\mathbf{C}(t),
  \label{eq:fci_coeff_eom}
\end{equation}
with the FCI ground state as the initial state.

Time propagation is performed using the operator-splitting technique of Ref.~\cite{krause2007molecular}.
When the external field is off, the propagation is analytic, $C_n(t+\Delta t)=C_n(t)e^{-iE_n\Delta t}$.
When the field is on, the coefficient vector is propagated according to
\begin{equation}
  \mathbf{C}(t+\Delta t)
  =
  \left[\prod_{q=x,y,z}
  \mathbf{U}_q^{\dagger}
  e^{iE_q(t)\tilde{\boldsymbol{\mu}}_q\Delta t}
  \mathbf{U}_q\right]
  e^{-i\mathbf{D}\Delta t}\mathbf{C}(t),
  \label{eq:fci_split_operator}
\end{equation}
where $\mathbf{D}_{mn}=E_m\delta_{mn}$ is the diagonal field-free FCI energy matrix, $E_q(t)$ is the $q$-component of the external electric field, and $\mathbf{U}_q$ diagonalizes the transition dipole matrix $\boldsymbol{\mu}_q$, i.e., $\tilde{\boldsymbol{\mu}}_q=\mathbf{U}_q\boldsymbol{\mu}_q\mathbf{U}_q^{\dagger}$ is diagonal.
This form exploits the fact that both the field-free propagation in the FCI eigenstate basis and the dipole kick in the dipole eigenbasis can be evaluated by diagonal matrix exponentials.

The grid-based RT-TDDFT reference calculation is performed for the stretched H$_2$ molecule using the \textsc{Octopus} package~\cite{castro2006octopus,andrade2015octopus}.
We use a cylindrical simulation box of radius $6.0\,\text{\AA}$ and length $12.0\,\text{\AA}$ parallel to the molecular axis, discretized with grid spacing $0.4$~a.u.
The Poisson equation is solved using the interpolating-scaling-function (ISF) solver.
Exchange--correlation effects are treated within the adiabatic approximation using the \texttt{gga\_x\_lb + lda\_c\_pz\_mod} functional.
To suppress spurious reflections in the weak-ionization regime, we employ a complex absorbing potential (CAP) with width $3.0\,\text{\AA}$ and height parameter $-1.0$.
Time propagation is carried out with the enforced time-reversal symmetry (ETRS) propagator using a time step $\Delta t=0.005$~a.u.

\section{Full Time Trajectories of H$_2$ Dipole Moment Evolution}
\begin{figure}[htbp]
  \centering
  \includegraphics[width=\textwidth]{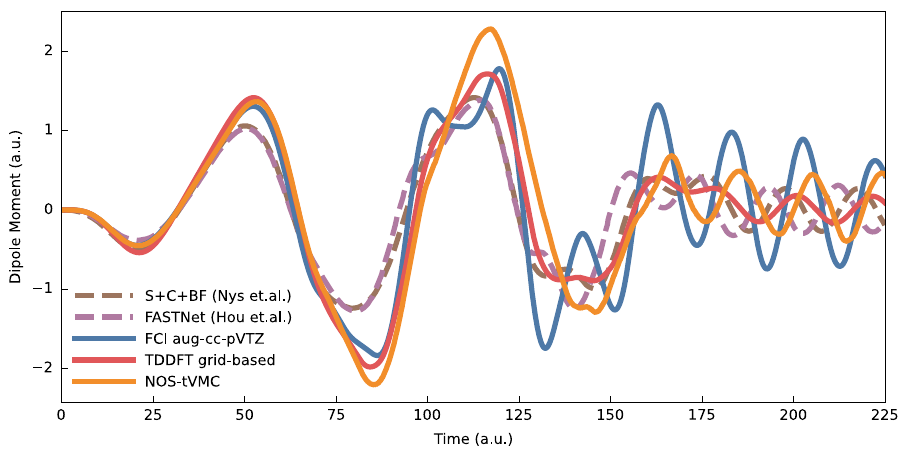}
  \caption{The full time trajectories of the dipole moment evolution of H$_2$ molecule under the electric field.}
  \label{fig:H2_full}
\end{figure}

\newpage
\section{Dynamic Polarizability Fitting Details}
\begin{figure}[htbp]
  \centering
  \includegraphics[width=\textwidth]{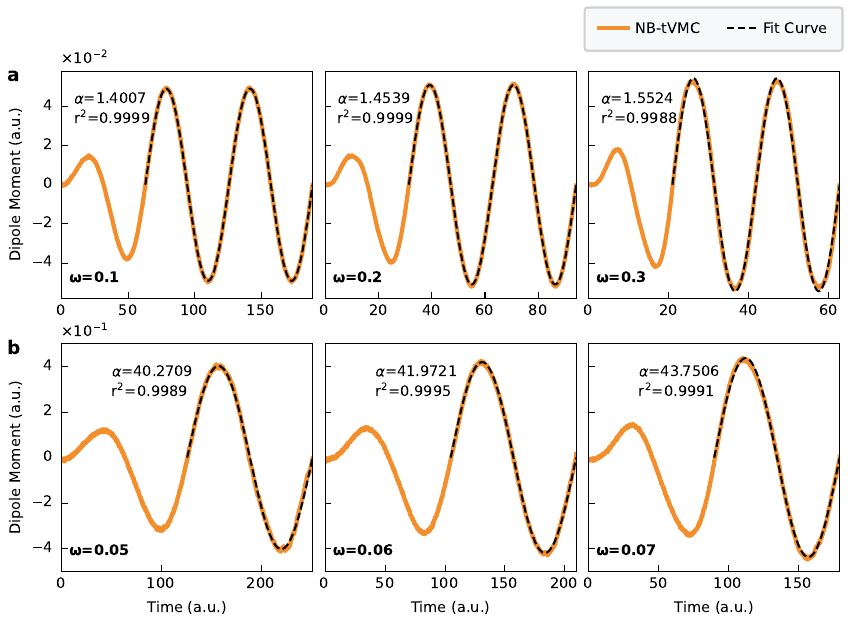}
  \caption{Dipole moment responses to single-frequency electric fields across different frequencies and the corresponding dynamic polarizability fitting results of (a) He atom and (b) Be atom. Simulating the Be atom presents greater challenges, resulting in unstable and less accurate evolution trajectories. Therefore, only the second optical cycle is extracted for the fitting process. The field amplitude $E_{\mathrm{max}}$ is 0.035~a.u. for the He atom and 0.01~a.u. for the Be atom.}
  \label{fig:dp_si}
\end{figure}

\nocite{*}
\bibliography{si}